\documentclass[aps,twocolumn,prc,superscriptaddress,showpacs,nofootinbib,amssymb,amsfonts,amsmath]{revtex4-1}

\usepackage{graphicx}
\usepackage{dcolumn}
\usepackage{bm}

\usepackage{amsmath}    
\usepackage{amsfonts}   
\usepackage{amssymb}
\usepackage{graphicx}   

\begin{document}

\title{Description of $^{158}$Er at ultrahigh spin in nuclear 
density functional theory.}

\author{A. V. Afanasjev}
\affiliation{Department of Physics and Astronomy, Mississippi State
University, Mississippi 39762, USA}
\affiliation{Joint Institute for Heavy-Ion Research, Oak Ridge, Tennessee
37831, USA}
\author{Yue Shi}
\affiliation{State Key Laboratory of Nuclear Physics and Technology,
School of Physics, Peking University, Beijing 100871, China}
\author{W.\ Nazarewicz}
\affiliation{Department of Physics and Astronomy, University of Tennessee,
Knoxville, Tennessee 37996, USA}
\affiliation{Physics Division, Oak Ridge National Laboratory, Oak Ridge, Tennessee 37831, USA}
\affiliation{Institute of Theoretical Physics, University of Warsaw, ul. Ho\.za 69,
PL-00-681 Warsaw, Poland }%

\date{\today}

\begin{abstract}
Rotational bands in $^{158}$Er at ultra-high spin  have
been studied in the framework of relativistic and
non-relativistic nuclear density functional theories.  Consistent results
are obtained across the theoretical models used but some puzzles remain when 
confronted with experiment. Namely,
the many-body
configurations which provide good description of experimental transition
quadrupole moments and dynamic moments of inertia require
substantial increase of the spins of observed bands as compared with
experimental estimates, which are still subject to large uncertainties. If, 
however,  the theoretical spins assignments turned out to be correct, the experimental
band 1 in $^{158}$Er would be the highest-spin structure ever observed.
\end{abstract}


\maketitle

The existence of nuclei with triaxial shape deformations has been a topic of 
active research since the early fifties \cite{Bohr52,Wheeler53}. So far,  
there has been no clear evidence for nuclei that are triaxial in their ground
 states, and theoretical mass table calculations predict very few such candidates
 having fairly small energy gain due to triaxiality \cite{Molgamma}. By far 
the clearest signatures come from the $\gamma$-ray spectroscopy of rotating 
nuclei since the angular momentum alignment of nucleons in  high-$j$ orbitals 
creates the shell structure which favors triaxiality at certain combinations 
of proton and neutron numbers, and specific rotational frequencies \cite{Szy83}. 
However, the evidence for existence of static triaxial shapes still remains scarce. 
It is generally accepted that the smooth terminating bands evolve gradually 
through the $\gamma$-deformation plane on approaching band termination \cite{PhysRep}; 
this feature is supported by the measurements of transition quadrupole moments 
\cite{Qsnsb}. The rotational bands associated  with the rigid triaxial shapes show 
specific features that allow to distinguish them from axially symmetric structures. 
Here, excellent examples are wobbling \cite{BM.75,J.02} and chiral \cite{F.01}  bands.

Triaxial superdeformed (TSD)  bands represent another class of structures built on 
static triaxial shapes. Of particular interest are the bands recently observed at 
ultrahigh spins in the $A\sim 154-160$ mass region. These are the bands seen in 
$^{154}$Er \cite{154Er-TSD}, $^{157,158,159,160}$Er 
\cite{Er158-PRL,Er158-Ruth,Er159-60.UH,Er159,160Er}, $^{157}$Ho \cite{Er158-Ruth}, 
and $^{160}$Yb \cite{Yb160}. As discussed below, the  interpretation of these 
structures is still under debate. The cranking calculations of Refs.\ \cite{BR.83,DN.85} 
predicted that collective triaxial configurations with large quadrupole deformations 
become competitive for spins above 50$\hbar$. More recent cranked Nilsson-Strutinsky 
(CNS) analysis of potential energy surfaces at the spins of interest have revealed
 the existence of three local minima \cite{Er-Qt}, namely, TSD1 with 
$\varepsilon_2 \sim 0.34$ and positive value of $\gamma \sim 20^{\circ}$, TSD2 with 
$\varepsilon_2 \sim 0.34$ and negative value of $\gamma \sim -20^{\circ}$, and TSD3 
with $\varepsilon_2 \sim 0.45$ and positive value of $\gamma \sim 25^{\circ}$. 
(For consistency with earlier publications, we adopt the CNS labeling of triaxial 
minima in this work. Note, however, that  DFT calculations yield  $\gamma-$deformations
 that  are typically smaller in absolute value than the ones obtained in  CNS.) The 
early interpretation of observed TSD bands have invoked configurations built either 
upon the TSD1 minimum \cite{154Er-TSD,Er158-PRL} or  TSD1 and/or TSD2 minima 
\cite{Er159-60.UH}.

However, the recent measurements of transition quadrupole moments of the TSD bands 
in $^{157,158}$Er nuclei \cite{Er-Qt} ruled out -- at least for these two nuclei --  
the interpretation based on the TSD1 minimum since the associated transition quadrupole 
moments of $Q_t\sim 7.5$ $e$b are significantly lower than the experimental values 
of $\sim 11$ $e$b. In addition, it was shown in the recent self-consistent tilted-axis 
cranking (TAC)  study \cite{SDFN.12} that the excited minimum TSD2 becomes a saddle 
point if the rotational axis is allowed to change direction, i.e.,  the mere appearance
 of TSD2 is an artifact of one-dimensional cranking approximation. The TAC work 
suggested the interpretation of  observed TSD bands in terms of TSD3. Such option has 
also been considered  in the CNS calculations  of Refs.~\cite{Er-Qt,KRMR.12}, however, 
no detailed study of it has been performed. On the contrary, a TSD2 scenario has been 
put forward in  the most recent paper \cite{KRMR.12}. The goal of this manuscript is 
to perform the detailed analysis of the observed TSD bands within the self-consistent 
framework based on nuclear density functional theory (DFT).

Our calculations for $^{158}$Er have been carried out  within two complementary 
theoretical methods; namely, relativistic (covariant) DFT \cite{VALR.05} and 
non-relativistic Skyrme-DFT \cite{BHR.03}. Since the TAC analysis \cite{SDFN.12} 
does not indicate the presence of tilted-axis solution for the TSD3 bands, we apply 
the principal axis cranking approximation. Moreover, because of very large angular 
momenta involved, pairing correlations are neglected. The resulting schemes are 
referred to as the cranked relativistic mean field (CRMF) \cite{CRMF} and cranked 
Skyrme Hartree-Fock (CSHF). In the CRMF calculations, all fermionic and bosonic 
states belonging to the shells up to $N_F = 14$ and $N_B = 20$ are taken into account 
in the diagonalization of the Dirac equation and the matrix inversion of the 
Klein-Gordon equations, respectively.  The NL1 \cite{NL1} and NL3* \cite{NL3*} 
parametrizations are used for the RMF Lagrangian; according to the recent study 
\cite{AS.11} these sets provide a reasonable description of the deformed 
single-particle energies. In the CSHF calculation, we use the symmetry unrestricted  
Hartree-Fock solver HFODD (v2.49) \cite{schunck12} and the SkM* energy density 
functional \cite{SkM}, which gives reasonable results in this mass region 
\cite{SDFN.12}. The CSHF calculation are performed in the stretched harmonic 
oscillator basis consisting of 680 basis states. The detailed investigations indicate 
that these truncation schemes provide a very reasonable numerical accuracy.

\begin{figure}[htb]
\includegraphics[width=0.9\columnwidth]{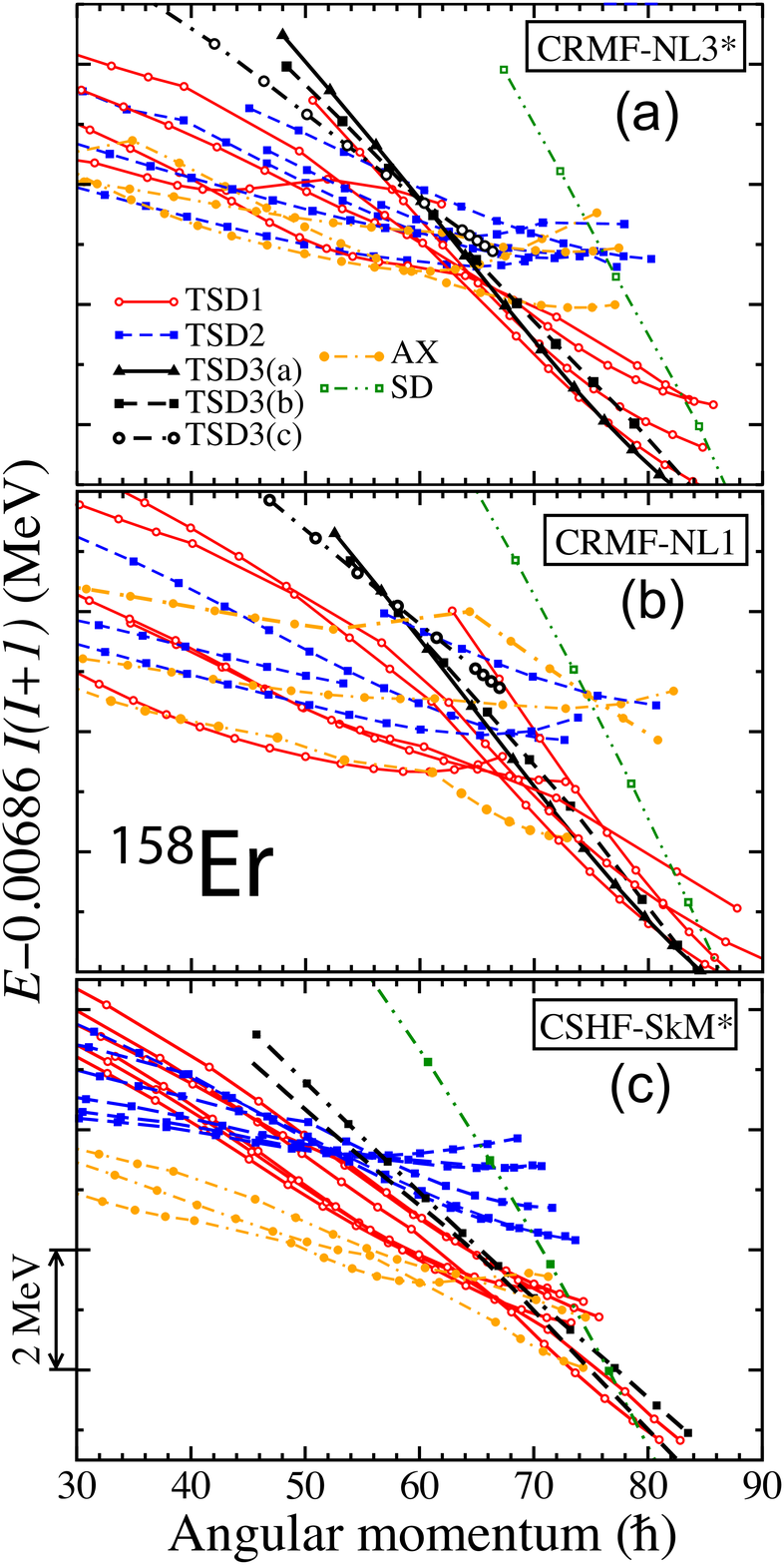}
\caption{(Color online) Energies of various configurations in $^{158}$Er 
calculated in the range of  $I=30-90$ using  (a) CRMF-NL3*, (b) CRMF-NL1, 
and (c) CSHF-SkM*,  relative to a smooth reference $E_{RLD}=AI(I + 1)$ 
with the inertia parameter $A = 0.00686$\,MeV. The lowest-energy 
near-prolate superdeformed configuration with quadrupole deformation   
$\beta_2 \geq 0.6$  is labeled as SD. See text for details.}
\label{E-E_RLD}
\end{figure}

The results of the calculations for the configurations at, or slightly above,  
the collective yrast line, are summarized in Fig.\ \ref{E-E_RLD}. The 
configurations can be divided into different groups according to their 
triaxiality and transition quadrupole moments in the spin range of 
$30-70\hbar$. Near-prolate configurations with $|\gamma|\leq 6^{\circ}$ are 
labeled AX. Their transition quadrupole moments are smaller than 9 $e$b so 
AX cannot be associated with the  observed bands. Triaxial configurations 
with positive $\gamma \sim 10^{\circ}$ and transition quadrupole moments 
$Q_t\leq 9$ $e$b are labeled  TSD1, while TSD2 are the configurations with 
negative $\gamma \sim -10^{\circ}$. Triaxial configurations with positive 
$\gamma \sim 13^{\circ}$ and transition quadrupole moments $Q_t\geq 9$ $e$b
 are labeled TSD3; their single-particle content is given in Table~\ref{tab1}. 
As mentioned earlier, $\gamma-$deformations in DFT are usually smaller 
than those in  CNS. 

\begin{table}[htb]
\caption{\label{tab1} TSD3 configurations studied in this work. Each 
configuration is defined by the number of neutrons and protons
occupying the four parity-signature $(\pi,r)$ blocks: $[(\pi=+,r=+i),(\pi=+,r=-i),
(\pi=-,r=+i),(\pi=-,r=-i)]$.}
\begin{ruledtabular}
\begin{tabular} {c c}
conf. label & configuration \\ \hline
TSD3(a) & $\nu [23,22,23,22]\otimes \pi [17,17,17,17]$ \\
TSD3(b) & $\nu [23,23,22,22]\otimes \pi [17,17,17,17]$ \\
TSD3(c) & $\nu [24,22,22,22]\otimes \pi [17,17,17,17]$ \\
TSD3(d) & $\nu [24,23,22,21]\otimes \pi [17,17,17,17]$ 
\end{tabular}
\end{ruledtabular}
\end{table}

It is rewarding to see that  the general structure of near-yrast bands 
predicted in $^{158}$Er weakly depends on the DFT model/parametrization 
used. However, the relative energies of the calculated configurations do 
depend on model/parametrization, thus reflecting the differences in the 
predicted energies of the single-particle states. Nevertheless, in all 
three models employed in our study, the configurations TSD3 approach 
yrast at $70\hbar$, and TSD2 always appear excited at very high spins 
(which rules them out \cite{SDFN.12}).  It is also important to mention 
that the configurations belonging to the same group have similar 
rotational inertia (slopes) in Fig.\ \ref{E-E_RLD}.


\begin{figure}[htb]
\includegraphics[width=\columnwidth]{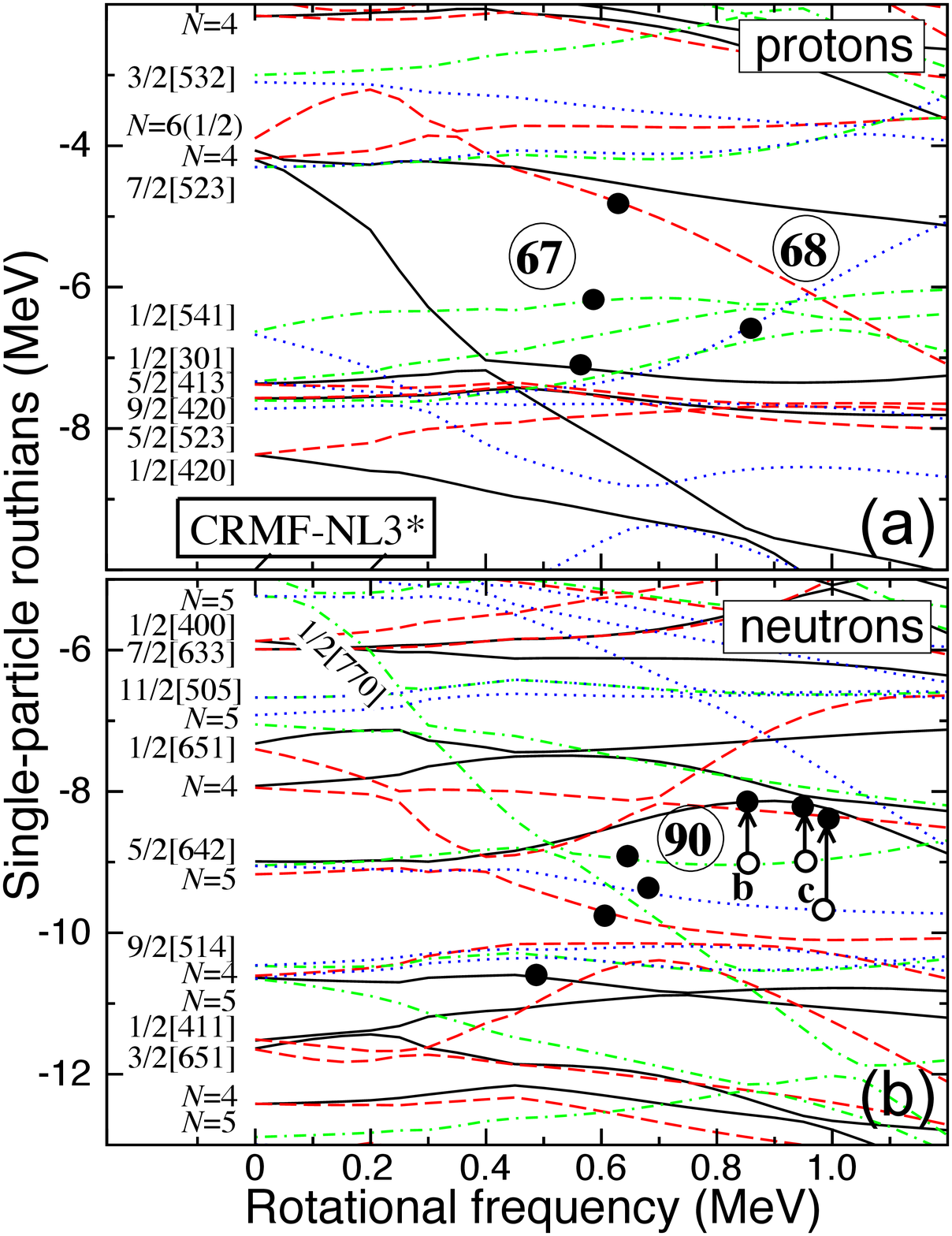}
\caption{(Color online) Proton (top) and neutron (bottom)
single-particle routhians  in CRMF-NL3* as a function of rotational frequency
$\Omega_x$.
They
are given along the deformation path of  TSD3(a)
at $\Omega_x \geq 0.45$ MeV; at lower frequencies the deformation
is fixed to the one obtained at $\Omega_x=0.45$ MeV.
Levels are labeled  by  parity $\pi$ and signature $r$ quantum numbers.
Solid,
short-dashed, dot-dashed and dotted lines indicate $(\pi=+, r=-i)$,
$(\pi=+, r=+i)$, $(\pi=-, r=+i)$ and $(\pi=-, r=-i)$ orbitals,
respectively. At $\Omega_x=0$ MeV,  single-particle routhians are
marked by means of either asymptotic quantum numbers $\Omega[Nn_z\Lambda]$
(in the case when the squared amplitude  of the dominant Nilsson  component
is greater than 0.5) or by the dominant principal quantum number $N$.
Solid circles indicate the last $(\pi,r)$ orbitals occupied within each 
$(\pi, r)$ family. Neutron particle-hole excitations leading to configurations
TSD3(b) and TSD3(c) are marked by arrows.}
\label{Routh}
\end{figure}

\begin{figure}[htb]
\includegraphics[width=0.9\columnwidth]{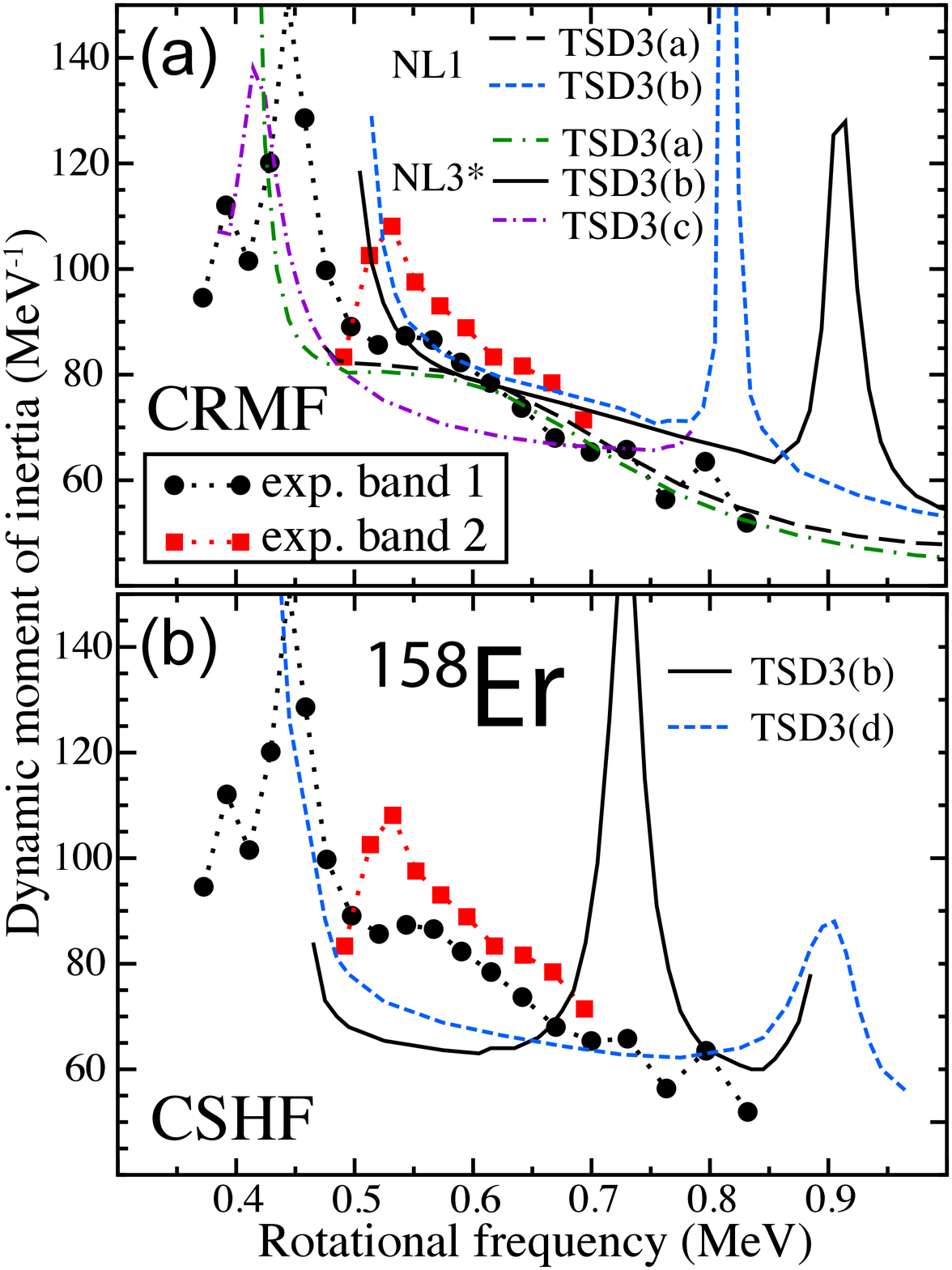}
\caption{(Color online) Experimental dynamic moments of inertia
of observed TSD bands in $^{158}$Er (symbols) compared to calculated 
ones (lines) in (a) CRMF and (b) CSHF. The calculated values are 
shown only in the frequency range in which the self-consistent 
solution corresponding to the band in question exist. See text for 
details.}
\label{J2-moments}
\end{figure}

The configuration TSD3(a) -- involving two protons in $N=6$ shell and 
one neutron in $N=7$ shell, i.e., $\pi 6^2 \nu 7^1$ --  is a possible 
CRMF candidate for the observed band 1. Single-particle CRMF-NL3* 
routhians corresponding to this configuration  are shown in Fig.\ 
\ref{Routh}. (Similar routhians have also been obtained in  CSHF.) 
In  CRMF, the transition quadrupole moment $Q_t$ of TSD3(a) changes 
from 10.5\,$e$b at $I=42$ to 9.0\,$e$b at $I=72$ and  $\gamma$ slightly 
increases  from $12^{\circ}$ to $16^{\circ}$ in this spin range. 
Considering that experimental value of $Q_t$ is subject to $\approx 15\%$ 
uncertainty due to nuclear and electronic stopping powers, these values 
are reasonably close to experiment. Moreover, as seen in  Fig.\ 
\ref{J2-moments}(a), the experimental dynamic moment of inertia $J^{(2)}$  
is rather well reproduced by assuming this configuration  above the band 
crossing at low frequencies; the level of agreement with experiment is 
comparable to  that earlier obtained for  superdeformed  bands in the 
$A\sim 150$ region \cite{CRMF,CRMF-align}.

Our CRMF-NL3* calculations suggest that the jump in dynamic moment of 
inertia of band 1 at low frequencies (see Fig.\ \ref{J2-moments}) can 
be associated with a band crossing with large interaction between the 
$1/2[770](r=+i)$ and $[N=5](r=+i)$ neutron routhians seen in  Fig.\ 
\ref{Routh}(b). In CRMF-NL1, the interaction between these routhians is 
weak so it can be removed by going to the diabatic representation 
\cite{PhysRep}.

\begin{figure}[htb]
\includegraphics[width=0.9\columnwidth]{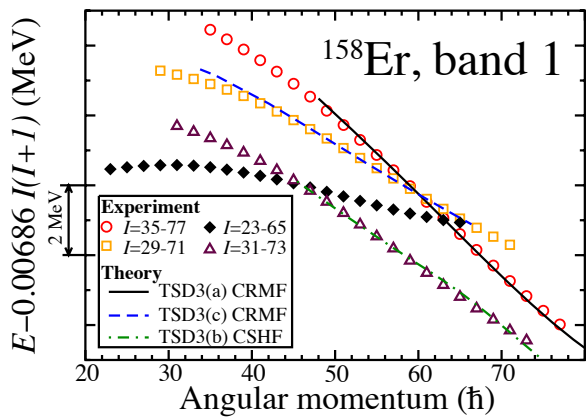}
\caption{(Color online) Similar as  in Fig.\ \ref{E-E_RLD} but
for experimental band 1 assuming different spin assignments (symbols), 
and for   calculated configurations TSD3(a) and TSD3(c) in  CRMF-NL3*,  
and TSD3(b) in CSHF-SkM* (lines). The energy of the lowest experimental state 
is selected arbitrarily to  minimize the deviation from calculated 
configurations.}
\label{E-E_RLD-exp}
\end{figure}

However, the interpretation of band 1 in terms of TSD3(a) is not 
consistent with the current experimental spin assignments.  A high-fold 
analysis of the intensity profiles at the bottom of band 1 in $^{158}$Er, 
compared to feeding intensities into the known yrast states, has allowed 
an estimation of the highest spin reached by this band to be 
$\sim 65\hbar$ \cite{Er158-PRL}.  At present, the uncertainty of this 
procedure is not obvious, however, the experimental error on spin 
assignment can be larger than $4\hbar$ \cite{R.priv}.  The comparison 
between experimental and calculated energies  shown in Fig.\ \ref{E-E_RLD-exp} 
indicates that, to be consistent with TSD3(a),  band 1 has to be observed
 in the spin range $I=35-77$. If this were the case, band 1 would be the 
rotational structure observed at the highest spin ever.  Considering the 
fact that this band carries only $\sim 10^{-4}$ of the respective channel 
intensity, i.e., two orders of magnitude less than the superdeformed yrast 
band in $^{152}$Dy  observed up to 68$\hbar$ \cite{Dy152}, this possibility 
cannot be excluded.

If  TSD3(a) is assigned to band 1, then the CRMF configuration TSD3(b) 
built upon TSD3(a) by exciting a neutron from $N=5$ $(r=+i)$ into  
$5/2[642](r=-i)$ is a natural candidate for experimental band 2. Indeed, 
as seen in Fig.\ \ref{J2-moments}, $J^{(2)}$  of this configuration is 
close to that of band 2, and its transition quadrupole moment is only 
slightly larger (by $\sim 0.7$ $e$b) than that of  TSD3(a).  Similar to 
the case of TSD3(a),  the increase of $J^{(2)}$ at low frequencies 
predicted for  TSD3(b) is due to an unpaired band crossing with a strong 
interaction between the $\nu 1/2[770](r=+i)$ and $N=5$ $(r=+i)$ orbitals. 
If this assignment is adopted for band 2, we must conclude that this 
structure is observed in a spin range $I=46-68$,  which again exceeds 
the experimental estimate \cite{Er158-PRL} by $8\hbar$.

According to  CRMF calculations,  an alternative configuration for band 1
  can also be suggested. This is the TSD3(c) configuration obtained from 
TSD3(a) by particle-hole excitations of two neutrons, see Fig.\ \ref{Routh}(b). 
 Although TSD3(c) somewhat underestimates experimental $J^{(2)}$, in a spin 
range of $I=34-66$ this configuration has  $Q_t=11-10.4$ $e$b  that is 
closer to experiment. In addition, the comparison of theoretical and 
experimental energies in Fig.\ \ref{E-E_RLD-exp} suggests that this 
configuration is consistent with the spin range of  $I=29-71$,  which 
is closer to experimental estimates. 

The problem with this interpretation is that TSD3(c) is never predicted 
very close to yrast (see Fig.\ \ref{E-E_RLD}). While the excitation energy 
of this configuration is relatively low in CRMF-NL3*, it is substantially 
larger in CRMF-NL1. It is interesting to note, however,  that shifting the 
neutron $N=5$ level -- from which the associated  particle-hole excitations 
are made --  by roughly 1 MeV in NL3*  results in a lowering  of TSD3(c) 
so that it becomes competitive  with TSD3(a) in the range of  $I=60-70$. 
The analysis of the deformed single-particle states at normal deformations 
\cite{AS.11} indicates that such a possibility cannot be excluded.

Similar results have also been obtained in our CSHF-SkM* calculations. 
Since the corresponding shell structure is slightly different as compared 
to CRMF-NL3* and CRMF-NL1, the energetics of  TSD3 bands predicted in CSHF 
is altered. For example, the lowest TSD3 configuration predicted in CSHF is 
TSD3(b); it was labeled  D  in Ref.\ \cite{SDFN.12}.  This configuration 
can be assigned to band 1. The configuration TSD3(a) is excited by $\sim$1 
MeV with  respect of TSD3(b). As seen in Fig.\ \ref{J2-moments}(b), the 
dynamic moment of inertia of  TSD3(b) above the low-frequency band crossing 
is  lower than that of experimental  band 1 and $J^{(2)}$ of CRMF TSD3(b). 
The latter maybe due to the particular choice of SkM* functional but the 
different treatment of time-odd mean fields in the CSHF and CRMF 
\cite{DD.95,AA.10} is also likely to  contribute. The dynamic moments of 
inertia of TSD3(b) in CSHF  shows an unpaired band crossing at $\Omega_x \sim 0.73$ 
MeV, which is absent in experiment. However, this should not prevent us from 
assigning  band 1 to TSD3(b) since the crossing frequencies strongly depend 
on relative energies of participating levels, which are not very accurately 
described in  DFT  \cite{SDMMNS.10,AF.05,AS.11}. A similar effect is also 
seen in the CRMF-NL3* and CRMF-NL1  results for  TSD3(b) in 
Fig.~\ref{J2-moments}(a), which predict a high-frequency unpaired band 
crossings at  frequencies which differ by as much as 0.1 MeV. As discussed 
in Ref.\ \cite{SDFN.12}, the calculated transition quadrupole moment $Q_t=10.7$ 
$e$b of TSD3(b) is close to experiment. The comparison between experimental 
and calculated energies in Fig.\ \ref{E-E_RLD-exp} indicates that band 1 
would have to be observed in the spin range of $I=31-73$ if TSD3(b) were 
assigned to this band in CSHF.

In  CSHF, band 2  is assigned to TSD3(d) -- a configuration built 
from TSD3(b) by promoting one neutron from the $[N=5](r=-i)$ 
routhian into  $1/2[651](r=+)$. The dynamic moment of inertia of 
this configuration is lower than the experimental one and the 
transition quadrupole moment of this configuration is $Q_t\sim 10.6$
 $e$b. The corresponding $\gamma-$deformation of this configuration 
is $\sim 11^{\circ}$. If this configuration assignment is adopted,  
band 2 should be observed in spin range $I=44-66$, which again is
 higher than experimental estimates of Ref.\ \cite{Er158-PRL}.

If experimental spin assignments ($I=23-65$) are used for band 1,
 then -- according to Fig.\ \ref{E-E_RLD-exp} -- only the  AX, TSD1, 
and TSD2 structures of Fig.\ \ref{E-E_RLD} can be considered as 
possible candidates for the observed band. However, the transition 
quadrupole moments predicted for AX and TSD1 are too small to 
explain experimental data. The results of Ref.\ \cite{SDFN.12} 
indicate that TSD2, which is predicted higher in energy in CSHF, 
is unphysical. This is consistent with the situation seen in 
Figs.~\ref{E-E_RLD}(b) and (c). However, in CRMF-NL3* variant of 
Fig.~\ref{E-E_RLD}(a) we find a TSD2 band that competes in energy 
with TSD1 up to $I \approx 62$. Here, a possibility that TSD2 represents 
a physical structure -- a scenario advocated in  Ref.~\cite{KRMR.12} -- 
cannot be ruled out.

In summary, properties of triaxial superdeformed  bands at ultra-high 
spin in $^{158}$Er have been analyzed in the framework of relativistic 
and non-relativistic DFT. The results obtained in these two theoretical 
frameworks are consistent. In particular, experimental transition 
quadrupole moments $Q_t$ and dynamic moments of inertia are well 
described by  TSD3 ($\pi 6^2 \nu 7^1$)  configurations having large 
quadrupole moments and positive $\gamma \approx 10-15^{\circ}$. The 
calculated spin assignments associated with these bands substantially 
exceed experimental estimates, which are still subject to large 
uncertainties. On the other hand, configurations which agree with 
experimental angular momentum assignments significantly  underestimate 
dynamic moments of inertia and transition quadrupole moments. If the 
theoretical spins assignments of Fig.\ \ref{E-E_RLD-exp} turned out to 
be correct, the experimental band 1 in $^{158}$Er would be the the 
highest-spin structure ever observed. The current study stresses the 
need for more precise  measurements of $Q_t$ and reliable estimates of 
spins in these bands.

  Useful discussions with J.\ Dobaczewski, I.\ Ragnarsson,  M.\ A.\ Riley, 
R.\ V.\ F.\ Janssens and F.\ Xu are gratefully acknowledged.  This work 
has been supported by the U.\ S.\ Department of Energy under Contracts 
Nos.\  DE-FG02-07ER41459 (Mississippi State University) and 
DE-FG02-96ER40963 (University of Tennessee), and by the Natural Science 
Foundation of China under Grant No.\ 10975006.

\end{document}